\documentclass[english,12pt,aps,prd,a4paper,preprintnumbers,floatfix,nofootinbib,showpacs,superscriptaddress, notitlepage]{revtex4-1} 
 \pdfoutput=1
\usepackage[usenames,dvipsnames]{color}  %%%%%%% usenames,dvipsnames required to get BrickRed \usepackage{color}
\usepackage{graphicx}
\usepackage{pifont}
\usepackage{caption}
\usepackage{subcaption}
\usepackage{amsmath}
\usepackage{amssymb}
\usepackage[colorlinks=true,citecolor=darkred,urlcolor=darkred, pdfborder={0 0 0}]{hyperref}
\usepackage[normalem]{ulem}

\usepackage[T1]{fontenc}
\usepackage[latin9]{inputenc}
\usepackage{array}
\usepackage{booktabs}
\usepackage{mathrsfs}
\usepackage{multirow}
\usepackage{tabularx}

\definecolor{darkred}{rgb}{0.6,0,0}

\definecolor{linkcolor}{rgb}{0,0,0.5}

\def\gsim{\raise0.3ex\hbox{$\;>$\kern-0.75em\raise-1.1ex\hbox{$\sim\;$}}}
\def\lsim{\raise0.3ex\hbox{$\;<$\kern-0.75em\raise-1.1ex\hbox{$\sim\;$}}}

\def\beqn#1{\begin{equation}\label{#1}}
\def\eeqn{\end{equation}}

\def\beqa#1{\begin{eqnarray}\label{#1}}
\def\eeqa{\end{eqnarray}}

\def\Z2{$\mathcal{Z_2}$}

\newcommand {\ignore}[1]{}

\newcommand{\sm}{{Standard Model }}

\def\321{$\mathrm{SU(3) \otimes SU(2) \otimes U(1)}$ }

\newcommand{\AddrAHEP}{%
  AHEP Group, Institut de F\'{i}sica Corpuscular --
  CSIC/Universitat de Val\`{e}ncia, Parc Cient\'ific de Paterna.\\
 C/ Catedr\'atico Jos\'e Beltr\'an, 2 E-46980 Paterna (Valencia) - SPAIN}

\newcommand{\AddrMiranda}{%
Departamento de F\'{\i}sica, Centro de Investigaci\'on
  y de Estudios Avanzados del IPN,\\ Apartado Postal 14-740 07000 Mexico,
  Distrito Federal, Mexico}

\newcommand{\AddrIoannina}{%
Division of Theoretical Physics, University of  Ioannina, GR 45110 Ioannina, Greece}

\begin{document}

\bibliographystyle{unsrt}   % needed for refs and hyperlinks %% utphys.bst style file is also needed with this %%%%%%%%%%%%

\title{\boldmath \color{BrickRed} Implications of the first detection of coherent elastic neutrino-nucleus scattering (CEvNS) with Liquid Argon}

\author{O. G. Miranda}\email{omr@fis.cinvestav.mx}
\affiliation{\AddrMiranda}

\author{D. K. Papoulias}\email{dimpap@cc.uoi.gr}
\affiliation{\AddrIoannina}

\author{G. Sanchez Garcia}\email{gsanchez@fis.cinvestav.mx}
\affiliation{\AddrMiranda}

\author{O. Sanders}\email{osanders@fis.cinvestav.mx}
\affiliation{\AddrMiranda}

\author{M. T\'ortola}\email{mariam@ific.uv.es}
\affiliation{\AddrAHEP}

\author{J. W. F. Valle}\email{valle@ific.uv.es}
\affiliation{\AddrAHEP}

\begin{abstract}
  
  The CENNS-10 experiment of the COHERENT collaboration has recently reported the first detection of coherent-elastic neutrino-nucleus scattering (CEvNS) in liquid Argon with more than $3 \sigma$ significance. In this work, we exploit the new data in order to probe various interesting parameters which are of key importance to CEvNS within and beyond the Standard Model.
  A dedicated statistical analysis of these data shows that the current constraints are significantly improved in most cases.
 We derive a first measurement of the neutron rms charge radius of Argon, and also an improved determination of the weak mixing angle in the low energy regime.
 We also update the constraints on  neutrino non-standard interactions, electromagnetic properties and light mediators with respect to those derived from the first COHERENT-CsI data.
\end{abstract}
\maketitle

\section{Introduction}

Over the past few decades, neutrino physics has entered the precision era and is steadily approaching a full description of the three neutrino oscillation picture~\cite{deSalas:2017kay}.
Although the next generation of long-baseline experiments such as DUNE and Hyper-Kamiokande will play a key role in settling the open issues, some degeneracies are likely to remain.
Moreover, beyond the challenges associated to the three-neutrino parameters, such as the atmospheric octant, the presence of leptonic CP violation, the neutrino mass ordering, and the absolute scale of the neutrino mass, there are a number of less standard, but important, neutrino properties to pin down.

The discovery of coherent elastic scattering of neutrinos off nuclei at the Spallation Neutron Source (SNS) at Oak Ridge National Laboratory (ORNL) using a CsI detector~\cite{Akimov:2017ade} has opened new ways to study key weak interaction parameters, such as the electroweak mixing angle~\cite{Cadeddu:2018izq,Canas:2018rng} and the nuclear form factors~\cite{Cadeddu:2017etk,Papoulias:2019lfi, AristizabalSierra:2019zmy}, as well as to probe novel neutrino properties beyond the Standard Model (SM)~\cite{Papoulias:2019xaw}. Among the most interesting possibilities, are those addressing non-standard interactions (NSI)~\cite{Liao:2017uzy,Dent:2017mpr,AristizabalSierra:2018eqm, Denton:2018xmq,Dutta:2019eml,Coloma:2017ncl,Gonzalez-Garcia:2018dep,Canas:2019fjw, Coloma:2019mbs, Dutta:2020che, Flores:2020lji}, electromagnetic neutrino properties~\cite{Kosmas:2015sqa,Kosmas:2017tsq,Cadeddu:2018dux,Miranda:2019wdy,Parada:2019gvy}, sterile neutrinos~\cite{Kosmas:2017zbh,Canas:2017umu,Blanco:2019vyp,Miranda:2019skf}, new light mediators~\cite{Farzan:2018gtr,Abdullah:2018ykz,Brdar:2018qqj, Billard:2018jnl, AristizabalSierra:2019ufd, AristizabalSierra:2019ykk,  Miranda:2020zji} and dark matter~\cite{Dutta:2019nbn, Akimov:2019xdj}.
The study of these scenarios would provide important hints for physics beyond the current three-massive-neutrino paradigm of elementary particles.
In fact, many of them may be regarded as implications of the very existence of neutrino masses themselves. 
Moreover, coherent elastic neutrino-nucleus scattering (CEvNS) experiments provide new ways to understand neutrino cross sections, crucial for establishing the robustness of oscillation experiments and their interpretation.

Here we show how the recent confirmation of the CEvNS process by the COHERENT collaboration using a 24~kg  liquid Argon (LAr) detector at the SNS~\cite{Akimov:2020pdx}, after collecting 6.12~GWh data substantially improves the sensitivity of a number of key weak interaction measurements. The reported results have shown a greater than $3 \sigma$ preference in favor of CEvNS. 

This paper is organized as follows. In Sec.~\ref{sec:basics} we set up the notation and summarize briefly the CEvNS formalism. Then, in Sec.~\ref{sec:analysis} we present the sensitivities on various parameters of the electroweak interaction in the SM  and beyond, obtained using the recent CENNS-10 results. Finally, our main conclusions are summarized in Sec.~\ref{sec:conclusions}.

\section{Basics}
\label{sec:basics}

Proposed more than forty years ago by Freedman~\cite{Freedman:1973yd},
this neutral current process is characterized by a cross section that increases as $N^2$, where $N$ being the number of neutrons in the nucleus:
\begin{equation}
\left(\frac{d \sigma}{dT_A}\right)_{\text{SM}} = \frac{G_F^2 m_A}{2 \pi}  (\mathcal{Q}^V_W)^2 \left[ 2 - \frac{2 T_A}{E_\nu} - \frac{m_A T_A}{E_\nu^2}\right]  \, ,
\label{eq:xsec-cevns}
\end{equation}
where $G_F$ denotes the Fermi constant, $T_A$ is the nucleus kinetic energy, $E_\nu$ the neutrino energy and
$\mathcal{Q}^V_W$ the vector weak charge written in the form
\begin{equation}
\mathcal{Q}^V_W =  \left[ g_{V}^{p} Z F_{p}(Q^{2}) +  g_{V}^{n} N F_{n}(Q^{2}) \right]  \, .
\end{equation}
Here, $Z$ and $N$ are the number of protons and neutrons in the
nucleus while the neutral current vector couplings, $g_{V}^{p,n}$, are
given by 
\begin{eqnarray}
g_{V}^{p} &=&\frac12-2 \sin^2 \theta_W \, ,
\nonumber 
\\ 
g_{V}^{n} &=&-\frac12 \, ,
\end{eqnarray}
with the weak mixing angle taken in the $\footnotesize{\overline{\text{MS}}}$ scheme, i.e.  $\sin^2 \theta_W \equiv \hat{s}^2_Z = 0.2312$. 
Finally, $F_{p,n}(Q)^{2}$ stands for the nuclear form factors for protons and neutrons respectively, for which we employ the well-known Helm parametrization
\begin{equation}
  F_{p,n}(Q^2) = 3\frac{j_1(QR_0)}{QR_0}\exp(-Q^2s^2/2) ,
  \end{equation}
where the magnitude of the three-momentum transfer is $Q=\sqrt{2m_AT_A}$, $j_1$ denotes the spherical Bessel function of order one and
$R_0^2=\frac53(R_{p,n}^2-3s^2)$ with $R_n=3.36$~fm ($R_p=3.14$~fm) denoting the neutron (proton) rms radius and $s=0.9$~fm.

To analyze the recent results of the liquid Argon detector reported by
the COHERENT collaboration, we will consider in this work a neutrino
flux arriving to the detector from the SNS at ORNL,  given by three different components from the
$\pi^+$ decay at rest. These are a ``prompt'' monoenergetic muon
neutrino signal given by
\begin{equation}
\frac{\mathrm{dN_{\nu _{\mu }}} }{\mathrm{d}E_\nu }=\eta\delta\left ( E_\nu-\frac{m_{\pi }^{2}-m_{\mu }^{2}}{2m_{\pi }} \right ) \, ,
\label{eq:PromptFlux}
\end{equation}
and the ``delayed'' neutrino flux composed of muon antineutrinos
\begin{equation}
\frac{\mathrm{dN_{\overline{\nu} _{\mu }}} }{\mathrm{d}E_\nu }= \eta \frac{64E_\nu^{2}}{m_{\mu }^{3}}\left ( \frac{3}{4}-\frac{E_\nu}{m_{\mu }} \right ) \, ,
\label{eq:DelFluxMuon}
\end{equation}
and electron neutrinos
\begin{equation}
\frac{\mathrm{dN_{\nu _{e }}} }{\mathrm{d}E_\nu }= \eta \frac{192E_\nu^{2}}{m_{\mu }^{3}}\left ( \frac{1}{2}-\frac{E_\nu}{m_{\mu }} \right ) \, , 
\label{eq:DelFluxEl}
\end{equation}
with the normalization factor $\eta$ given by $\eta = rN_{\text{POT}}/4\pi L^{2}$. The number of protons on target (POT) corresponding to the 6.12~GWh exposure is  $N_{\text{POT}}= 1.37\times10^{23}$,  $r=0.08$ denotes the produced neutrinos per POT and $L=27.5$~m is the CENNS-10 baseline.

In the present work, prompted by the importance of the liquid Argon detector of the COHERENT
collaboration, we study  relevant implications of
the measurement for different standard and new physics scenarios. 

\section{The analysis}
\label{sec:analysis}
\begin{figure}[t]
\includegraphics[width= 0.5\textwidth]{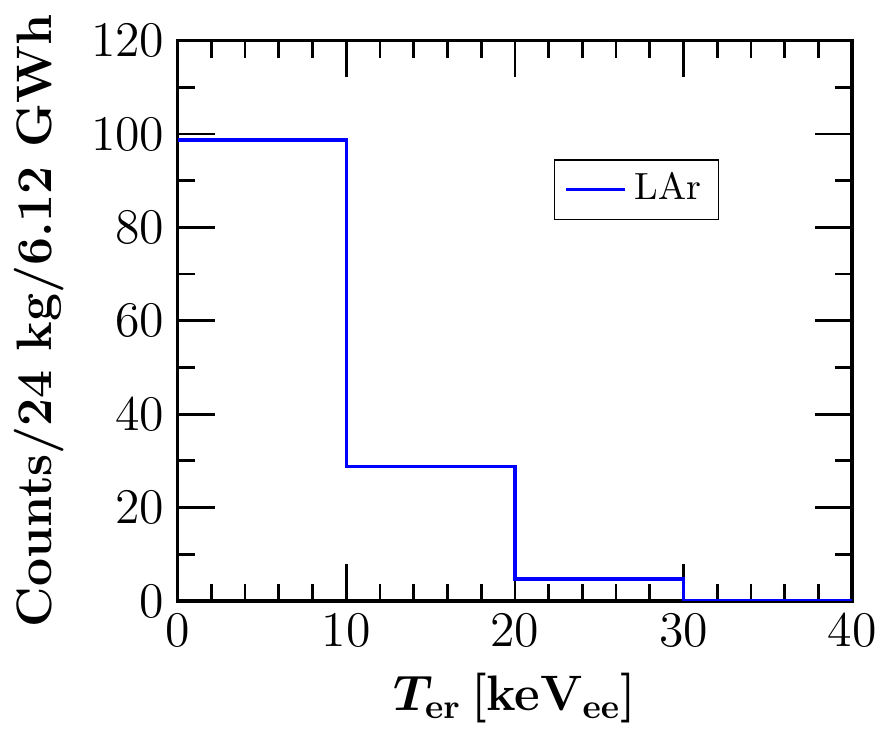}
\caption{Simulated number of events at the CENNS-10 LAr detector as a function of the electron recoil energy $T_{\mathrm{er}}$.}
\label{fig:events}
\end{figure}

As a first step, we simulate the CEvNS signal at the LAr detector using the efficiency function~\footnote{Taken from Fig.~3 of Ref.~\cite{Akimov:2020pdx}.} $\mathcal{A}(T_A)$, corresponding to the ``Analysis A''  of Ref.~\cite{Akimov:2020pdx}. We then convert the nuclear recoil spectrum into  electron recoil space through the reported quenching factor $\text{QF}(T_A)$. Following Ref.~\cite{Akimov:2020pdx}, we evaluate the number of events in the region of interest, i.e. for recoil energies below $120~\mathrm{keV_{nr}} \approx 30~\mathrm{keV_{ee}}$. Our simulated energy spectrum is shown in Fig.~\ref{fig:events} as a function of the electron recoil energy,  $T_\mathrm{er}$. 

Given the reliability of our simulated CENNS-10 signal, we perform  a statistical analysis of the recent liquid Argon
CEvNS measurement with the ultimate goal of probing important observables such as the neutron mean radius, the  electroweak mixing angle $\sin^2\theta_W$,
as well as to constrain new physics parameters. Before displaying our results in the next sections, here we
discuss the general procedure we followed to investigate the different scenarios. 

To test the sensitivity of the experiment to  observables under study, we
have performed a $\chi^{2}$ analysis, minimizing the function:
\begin{equation}
\chi^2 (X) =  \underset{\alpha}{\mathrm{min}} \Bigg [ \frac{\left(N_{\mathrm{meas}} - N_{\mathrm{theor}}(X) [1+ \alpha]  \right)^2}{\sigma_{\mathrm{stat}}^2} 
  + \left(\frac{\alpha}{\sigma_{\alpha}} \right)^2  \Bigg ] \, ,
\label{chi_sq_SNS}
\end{equation}
with $N_{\mathrm{meas}}=159$ denoting the number of the measured events from the fit of ``Analysis A'' in Ref.~\cite{Akimov:2020pdx} and
$N_{\text{theor}}(X)$ being the theoretical prediction. Here, the argument $X$ represents
the set of parameters to be tested, such as the weak mixing angle, the neutron rms radius or NSI parameters.
Here we note that we have successfuly calibrated our procedure with the one discussed in the recent CENNS-10 result.
In Eq.(\ref{chi_sq_SNS}), the statistical uncertainty is given by
   $\sigma_{\mathrm{stat}}=\sqrt{N_{\mathrm{meas}}+ N_{\mathrm{BRN}}}$, where
$N_{\mathrm{BRN}}=563$ represents the number of background events due to beam related neutrons (BRN).
The parameter $\alpha$ quantifies the normalization, that has a
systematic error $\sigma_{\alpha} =8.5\%$ (for details, see Ref.~\cite{Akimov:2020pdx}).

\subsection{Standard electroweak and nuclear physics}
\label{sec:SM}

\begin{figure}[t]
\includegraphics[width= 0.48 \textwidth]{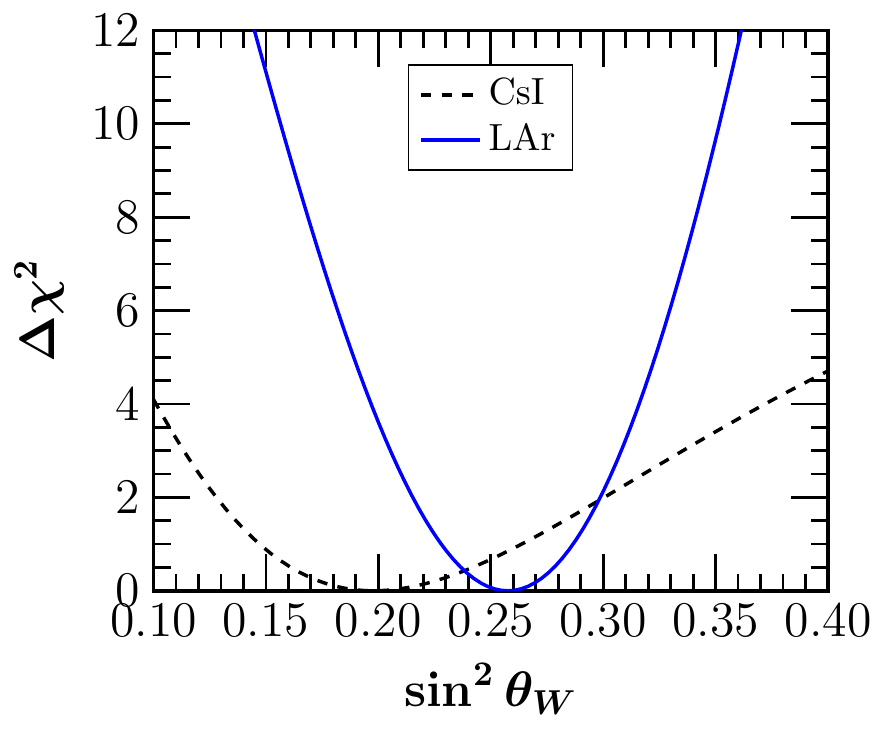}
\includegraphics[width =0.48 \textwidth]{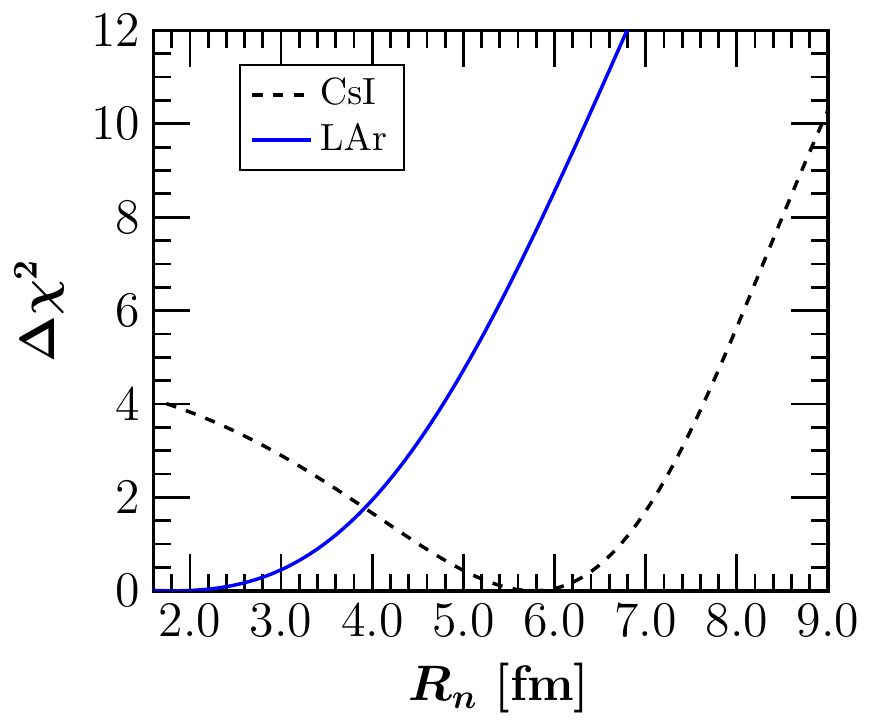}
\caption{Sensitivity on the weak mixing angle (left) and on the neutron rms radius (right). A comparison between the results obtained with LAr and CsI detectors is also shown.}
\label{fig:EW}
\end{figure}

One of the most important parameters of the SM is the weak mixing angle, that is measured with great accuracy at the $Z$~peak. At low energies, however, the existing measurements are less precise but still very relevant given the prediction of an increase of about 3\% in its value due to radiative corrections. Moreover, any deviation from 
the SM prediction for this value would be a signature of new physics. 
With this motivation in mind, we have performed a $\chi^2$ analysis for this SM parameter using the liquid Argon data.
Our corresponding results are shown in the left panel of Fig.~\ref{fig:EW}, where one can see a notable improvement with respect to the previous determination of $\sin^2 \theta_W$
from the COHERENT-CsI data (see Ref.~\cite{Kosmas:2017tsq}). The new measurement of the weak mixing angle, derived from the CENNS-10 data at 90\% C.L. reads

\begin{equation}
\sin^2 \theta_W = 0.258^{+0.048}_{-0.050}
\label{eq:sin2w-limit}
\end{equation}

Another very useful standard information that can be obtained from the CEvNS interaction is the neutron mean radius $R_n$ for the Argon isotope.
Although there are theoretical predictions for this value, its direct determination can facilitate a better understanding of the CEvNS background at dark matter oriented experiments~\cite{Papoulias:2018uzy}.
We have performed the corresponding analysis for this observable and the obtained result is illustrated in the right panel of Fig.~\ref{fig:EW}. Although not directly comparable, the average neutron rms radius of CsI is also shown for the reader's convenience. On the other hand, it is worthwhile to notice that  $R_n$ is more severely constrained for the case of Argon. The obtained 90\% C.L. limit for the neutron rms radius for Argon reads
\begin{equation}
R_n < 4.33 ~\mathrm{fm} \, .
\label{eq:Rn-limit}
\end{equation} 
We should note that this provides the first experimental determination of the neutron radius in Argon. In addition, comparing with the previous result on CsI~\cite{Cadeddu:2017etk} (see also Refs.~\cite{Papoulias:2019lfi,Papoulias:2019txv}), one sees that the level of precision seems somewhat improved.

\subsection{Non-standard Interactions}

\begin{figure}[!t]
\includegraphics[width= 0.47 \textwidth]{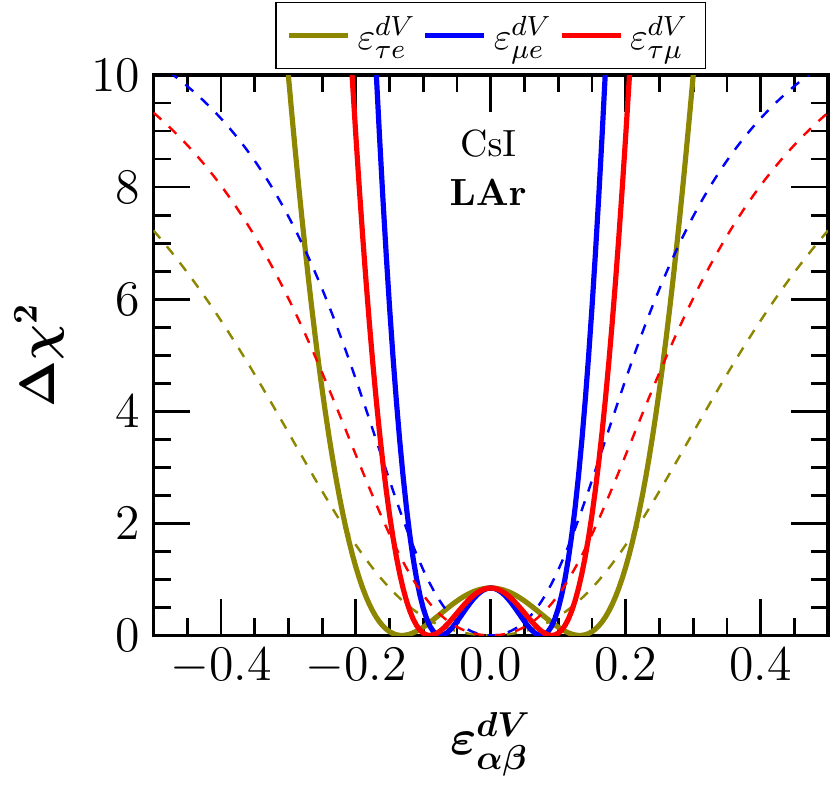}
\includegraphics[width= 0.5 \textwidth]{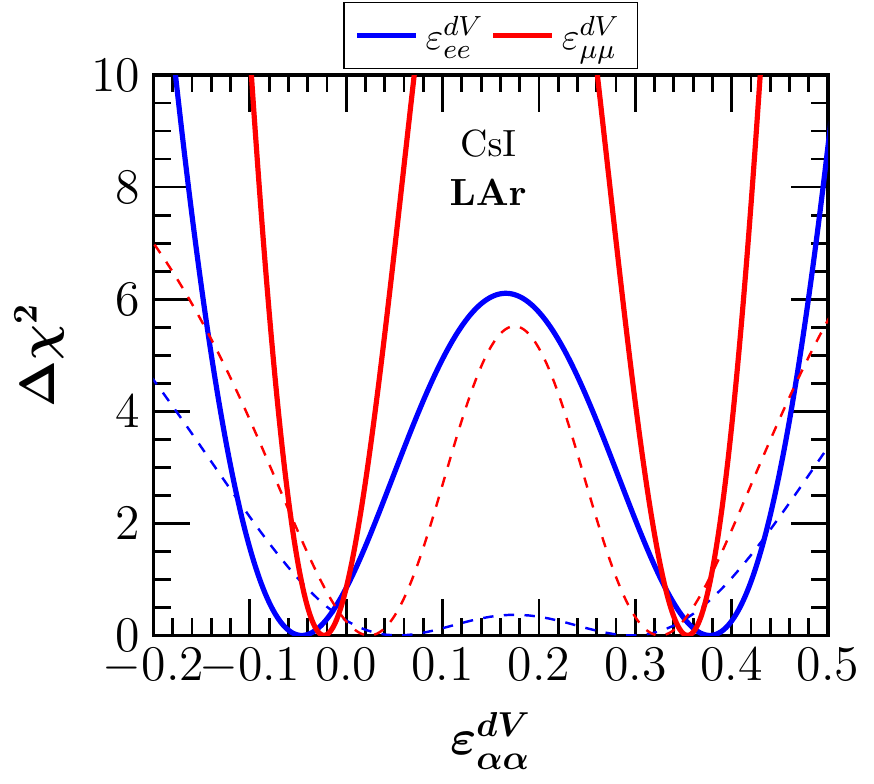}
\caption{Sensitivity of the new  CENNS-10 LAr results (solid lines) and the first  COHERENT-CsI measurements (dashed lines) to the flavor changing (left) and non-universal (right) NSI couplings. }
\label{fig:NSI}
\end{figure}
Most models that try to explain the neutrino mass pattern predict a modification of the $V-A$ couplings predicted by the SM. In many cases,
the corrections are expected to be negligible, like in the most simple type I seesaw model while, in other cases, there could be a relatively
larger signal, as for the linear and inverse seesaw cases. A large family of new physics models can be phenomenologically described  using
the formalism of NSI, that modify the neutral current SM Lagrangian through the contribution~\cite{Farzan:2017xzy,Miranda:2015dra,Ohlsson:2012kf}
\begin{equation}
{\cal{L}}^{NSI}_{NC}=-2\sqrt{2}G_F \sum\limits_{f,P,\alpha,\beta} \varepsilon_{\alpha\beta}^{f P}(\bar{\nu}_\alpha\gamma^\mu P_L\nu_\beta)(\bar{f}\gamma_\mu P_X f) \, ,
\end{equation}
where $f$ corresponds to an elementary fermion that, in the case of CEvNS, reduces to the quarks of the first family,
  $f=\{u,d\}$. $\alpha$ and $\beta$ denote the neutrino flavors $\{e,  \mu, \tau\}$, $P_X$ the left and right chirality projectors
  $P_{L,R}$, and $\varepsilon_{\alpha\beta}^{f P}$ are the couplings that quantify the relative strength of the NSI.
Due to the presence of these new interactions, the weak charge of the CEvNS reaction is modified according to the substitution
$\mathcal{Q}^V_W \to \mathcal{Q}^V_\text{NSI}$ in Eq.~(\ref{eq:xsec-cevns}), with the NSI charge given by
\begin{eqnarray}
\mathcal{Q}^V_\text{NSI} &=& 
  \left [ \left ( g_{V}^{p}+2\varepsilon _{\alpha \alpha}^{uV}+\varepsilon _{\alpha \alpha}^{dV}  \right )Z F_{p}(Q^{2})+\left ( g_{V}^{n}+\varepsilon _{\alpha \alpha}^{uV}+2\varepsilon _{\alpha \alpha}^{dV}  \right ) N  F_{n}(Q^{2})\right ]  \nonumber \\
  &+&\sum_{\alpha }\left [ \left ( 2\varepsilon _{\alpha \beta}^{uV}+\varepsilon _{\alpha \beta}^{dV} \right ) Z F_{p}(Q^{2})+ \left ( \varepsilon _{\alpha \beta}^{uV}+2\varepsilon _{\alpha \beta}^{dV} \right )N F_{n}(Q^{2})    \right ] .
\label{eq:qvNSI}
\end{eqnarray}

As it has already been noticed~\cite{Barranco:2005yy}, CEvNS reaction is sensitive to the NSI parameters and, therefore, it can provide
important information to probe the so-called LMA-Dark solution~\cite{Miranda:2004nb}. The results of our $\chi^2$ analysis, for one
NSI parameter at a time, are shown in Fig.~\ref{fig:NSI}, both for the flavor changing (left panel) and the non-universal case (right panel).
We can see from this figure that the sensitivities on the flavor changing NSI parameters are only marginally improved with respect to the previous CsI
  case. This is due to the detection of a larger number of events with respect to the SM prediction. Although the excess is below one standard deviation, still
  the preferred value for the flavor changing parameters are non-zero, as can be seen in Fig.~\ref{fig:NSI}. In the case of non-universal NSI, one can see the improvement in their restriction
in comparison with the first CEvNS detection. \\[-.2cm]

\begin{figure}[t]
\includegraphics[width= 0.48 \textwidth]{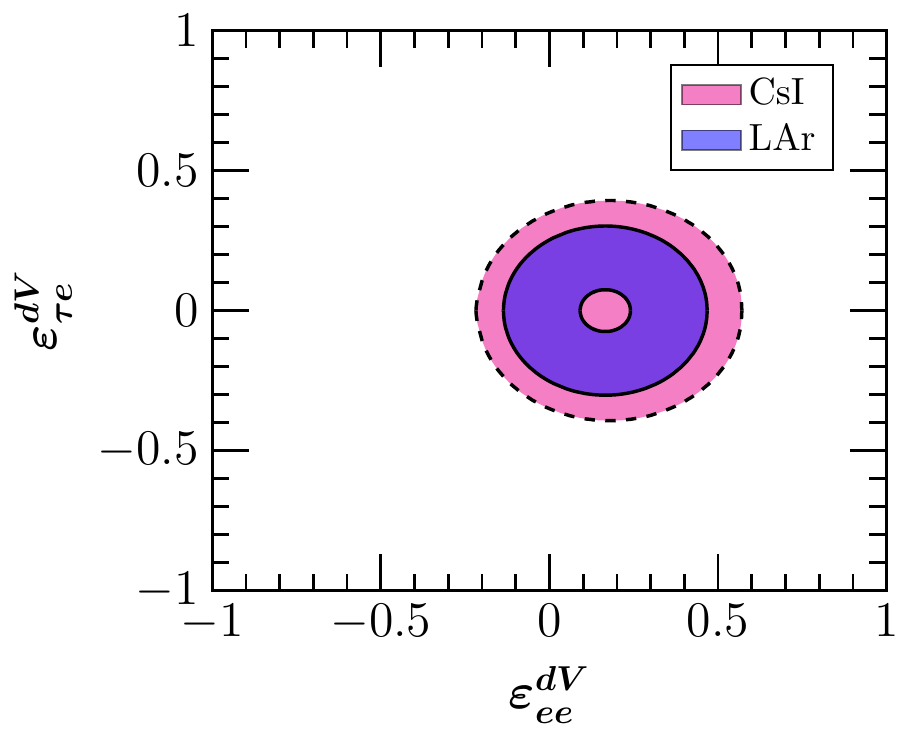}
\includegraphics[width= 0.48 \textwidth]{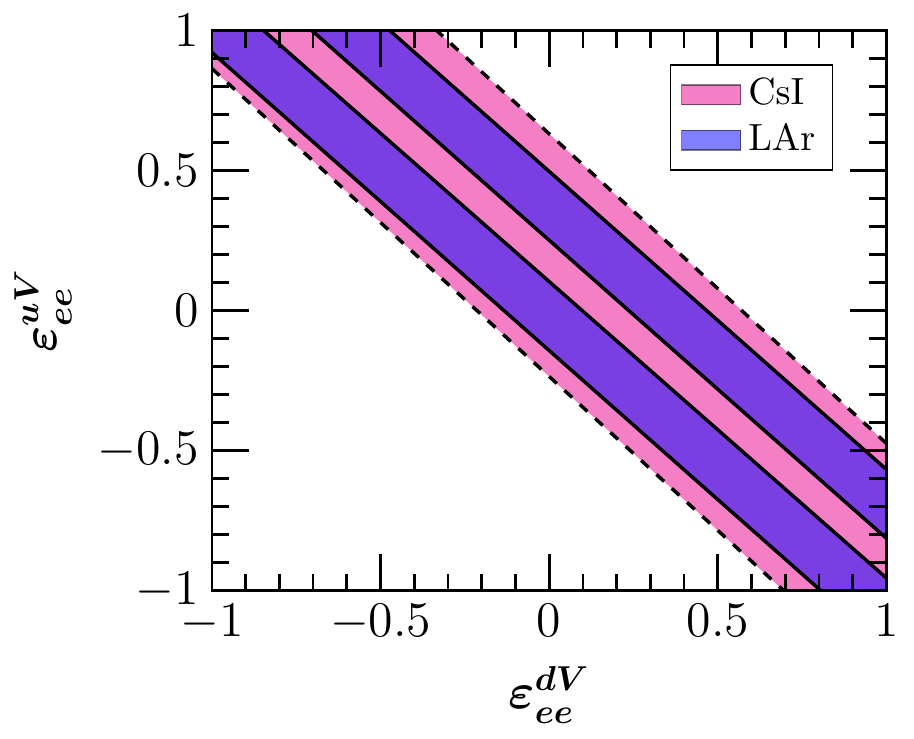}
\caption{ 90\% C.L. allowed regions from the analysis allowing two NSI parameters at a time. The left panel considers the simultaneous presence of non-universal and flavor-changing NSI with $d$ quark, while the right panel  corresponds to the case of simultaneous non-universal NSI couplings with $u$ and $d$ quarks. For comparison, we show the results from the analysis of CsI and LAr data.}
\label{fig:NSI-2par}
\end{figure}

 We can go one step further in the analysis and study more general restrictions on NSI parameters~\cite{Barranco:2005yy,Scholberg:2005qs}.
  For example, the new CENNS-10 measurement can be used to constrain pairs of NSI parameters, allowing us to seek for possible correlations between them, as shown in Fig.~\ref{fig:NSI-2par},
  where the constraints on the two-dimensional parameter space of non-universal and flavor-changing NSI couplings with $d$ quark ($\varepsilon_{ee}^{dV}$, $\varepsilon_{\tau e}^{dV}$) are given (left panel),
  as well as those for the case of the non-universal NSI couplings with $d$ and $u$ quarks ($\varepsilon_{ee}^{dV}$, $\varepsilon_{ee}^{uV}$). In both cases one can appreciate the improvement in the determination of the parameters. The good consistency of our analysis with the available results of Ref.~\cite{Akimov:2020pdx} is evident from the right panel of this figure.

\subsection{Electromagnetic properties}

The discovery of neutrino oscillations with solar, atmospheric, reactor  and accelerator neutrinos~\cite{deSalas:2017kay} has set a new milestone in particle physics pointing to the existence of massive neutrinos, hence constituting the clearest signature of new physics beyond the SM. The non-vanishing neutrino mass stands out as the best motivation for non-trivial electromagnetic (EM) neutrino properties. Indeed, the expansion of the EM neutrino vertex (for details see Ref.~\cite{Giunti:2014ixa}) yields two main phenomenological parameters, namely the neutrino magnetic moment and the neutrino charge radius~\cite{Vogel:1989iv}.

\begin{figure}[t]
\includegraphics[width= 0.48 \textwidth]{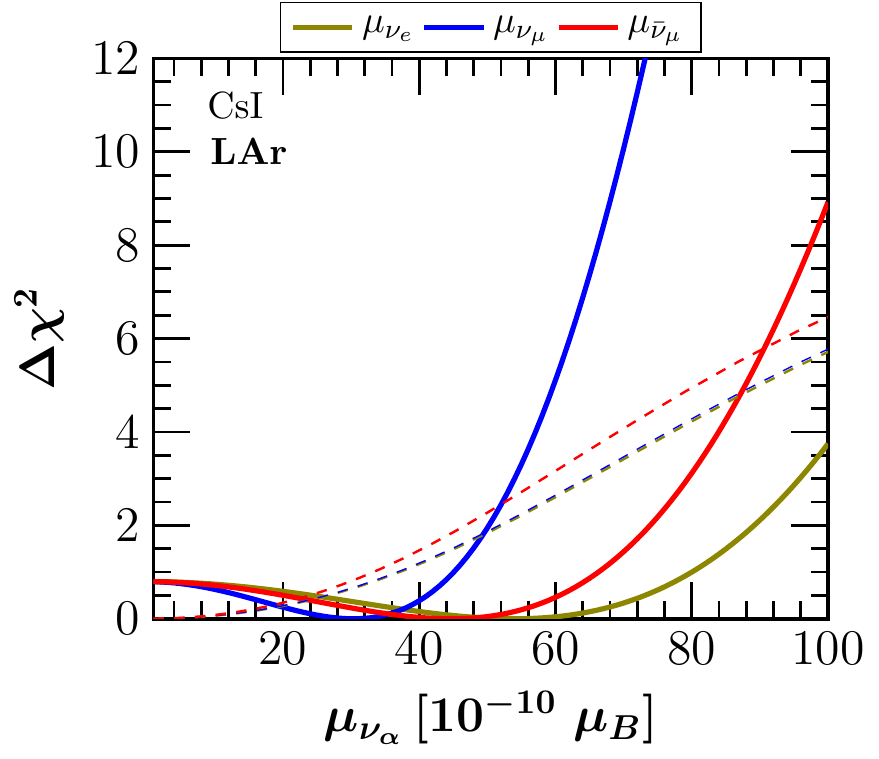}
\includegraphics[width= 0.48 \textwidth]{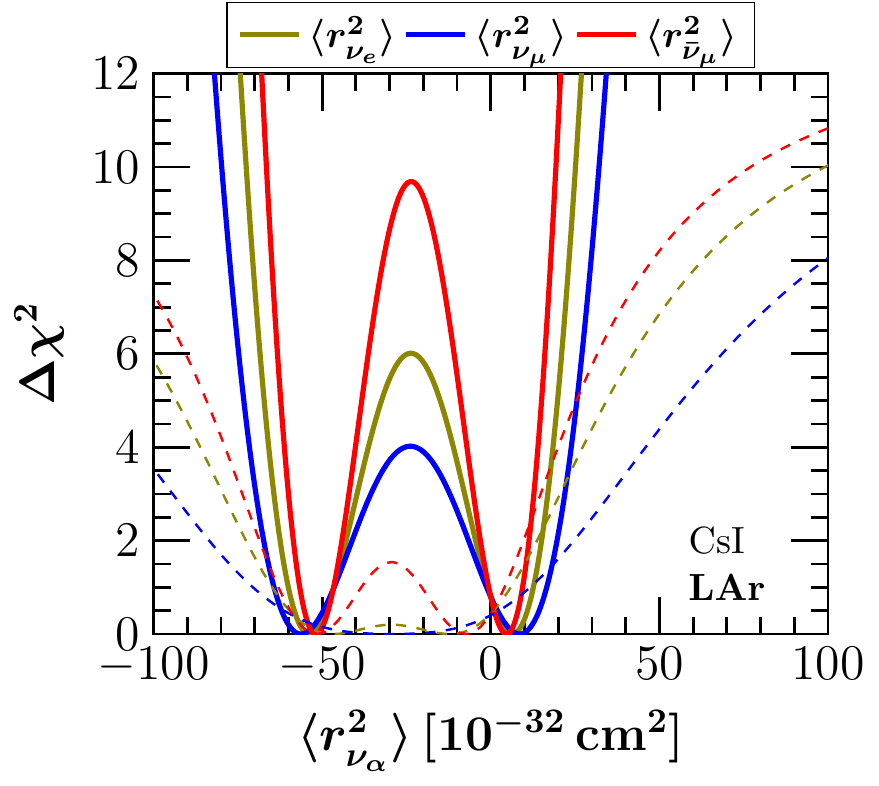}
\caption{Sensitivity to the neutrino magnetic moment (left) and charge radius (right). Thick (thin) curves correspond to the LAr (CsI) measurement.}
\label{fig:EM}
\end{figure}

The neutrino magnetic moment $\mu_{\nu_\alpha}$ with $\alpha= e, \mu, \tau$ is a flavor-dependent quantity  which, for the case of scattering experiments, is usually expressed in the mass basis~\cite{Grimus:2002vb,Tortola:2004vh}.  Due to the helicity-violating nature of the EM cross section, there is no interference with the SM one given in Eq.(\ref{eq:xsec-cevns}) and yields an additive contribution of the form
\begin{equation}
\left( \frac{d \sigma}{dT_A} \right)_{\mathrm{EM}}=\frac{\pi a^2_{\text{EM}} \mu_{\nu}^{2}\,Z^{2}}{m_{e}^{2}}\left(\frac{1-T_A/E_{\nu}}{T_A}\right) F_p^{2}(Q^{2})\,,
\label{NMM-cross section}
\end{equation}
where the flavor index has been dropped. For sufficient low detection threshold, the latter cross section leads to an enhancement of the recoil spectrum, i.e. a feature that is not expected for the other types of new physics considered in this work. Note that, for the case of Majorana neutrinos, only transition magnetic moments are expected, and the corresponding sensitivities from the analysis of neutrino-electron scattering~\cite{Canas:2015yoa} and CEvNS~\cite{Miranda:2019wdy} experiments have already been given in the literature. Here, for simplicity, we only consider flavor-dependent effective neutrino magnetic moments. The resulted sensitivity profiles relevant to $\mu_{\nu_e}$, $\mu_{\nu_\mu}$ and $\mu_{\bar{\nu}_\mu}$ from the analysis of the recent CENNS-10 data are shown in the left panel of Fig.~\ref{fig:EM}. The obtained constraints at 90\% C.L. read 
\begin{equation}
\left( \mu_{\nu_e}, \mu_{\nu_\mu}, \mu_{\bar{\nu}_\mu} \right) < (94, 53, 78)~10^{-10}\mu_B\, .
\end{equation}
We notice that the LAr sensitivity on the effective neutrino magnetic moment is not drastically improved in comparison to the one reached with the first COHERENT-CsI data~\cite{Kosmas:2017tsq}.
  This is due to the fact that the CENNS-10 result of $N_{\text{meas}}= 159$ events is by about 30 events larger than the 130~events expected in the Standard Model.
  Therefore, even though the systematic uncertainties of CENNS-10 experiment are better compared to the first CsI measurement, these extra 30 events in the $\chi^2$ translate into a finite neutrino magnetic moment contribution, thus weakening the limits.

\begin{figure}[t]
\includegraphics[width= \textwidth]{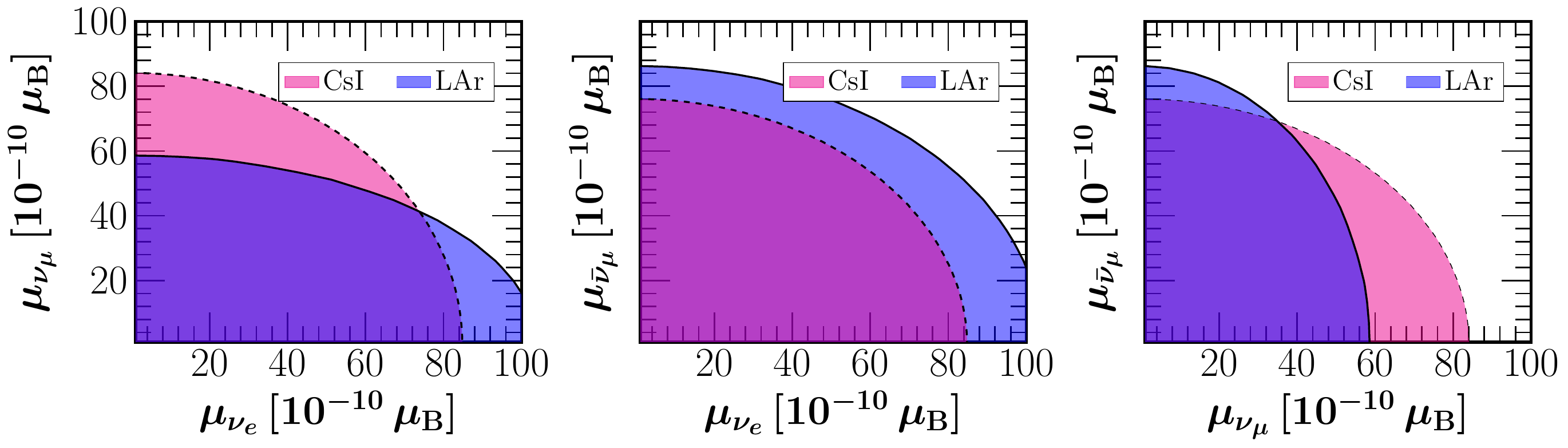}
\includegraphics[width= \textwidth]{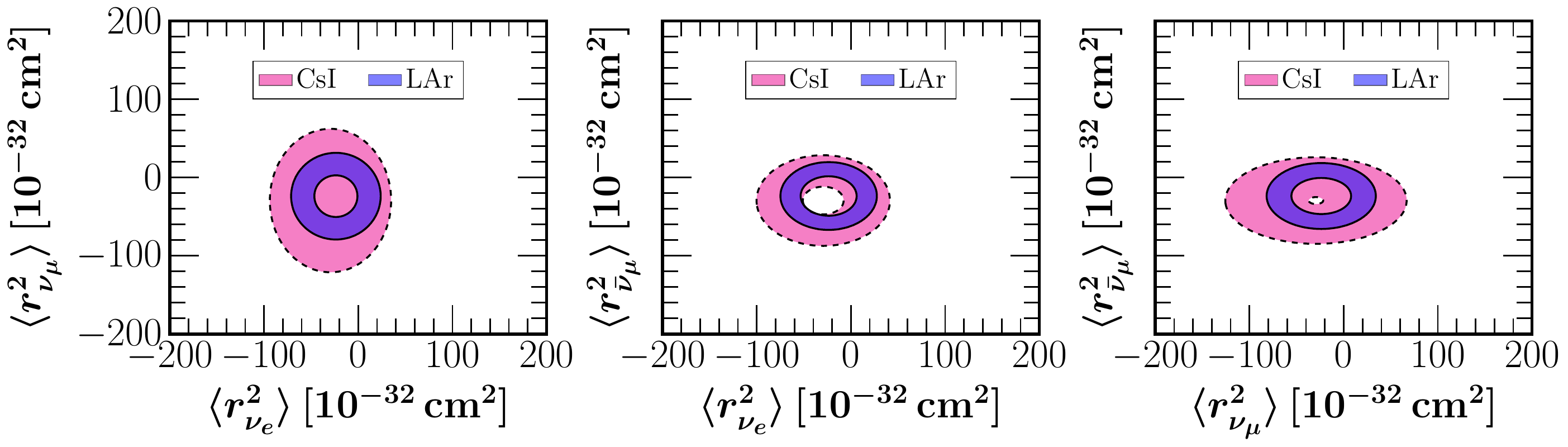}
\caption{Upper panel: 90\% C.L. allowed region in the parameter space of the neutrino magnetic moments $(\mu_{\nu_\alpha}, \mu_{\nu_\beta})$.  Lower panel: 90\% C.L. allowed region in the parameter space of neutrino charge radii $(\langle r^2_{\nu_\alpha}\rangle, \langle r^2_{\nu_\beta}\rangle)$. The results are shown for different choices of neutrino flavours, with the undisplayed parameters in each case assumed to be vanishing. For comparison, we show the results from the analysis of CsI and LAr data.}
\label{fig:comb-EM}
\end{figure}

We now turn our attention to the neutrino charge radius $\langle r_{\nu_\alpha}^2\rangle$ with $\alpha= e, \mu, \tau$ being the flavor index. Likewise the neutrino magnetic moment, $\langle r_{\nu_\alpha}^2\rangle$ is also required to be expressed in the mass basis through a rotation with the lepton mixing matrix~\cite{Cadeddu:2018dux}. Being a helicity-preserving quantity, its impact to the SM cross section is simply taken as a shift on the weak mixing angle according to 
\begin{equation}
\sin^2 \theta_W \rightarrow \hat{s}^2_Z + \frac{\sqrt{2} \pi \alpha_{\mathrm{EM}}}{3 G_F} \langle r_{\nu_\alpha}^2\rangle \, .
\label{eq:rv}
\end{equation}
Note that only the proton coupling $g_V^p$, proportional to the number of protons, interacts with the charge radius\footnote{We neglect the contribution of the term proportional to $Z^2$ corresponding to transition charge radii~\cite{Cadeddu:2018dux}.}. On the other hand, when antineutrinos are involved, both the $g_V^p$ and $\langle r_{\nu_\alpha}^2\rangle$ change sign and, therefore, Eq.(\ref{eq:rv}) holds for both neutrinos and antineutrinos, a result that is consistent with Ref.~\cite{Kosmas:2017tsq}. From our analysis of the recent CENNS-10 data, we obtain the sensitivity to the neutrino charge radii corresponding to $\langle r_{\nu_e}^2\rangle$, $\langle r_{\nu_\mu}^2\rangle$ and $\langle r_{\bar{\nu}_\mu}^2\rangle$, as shown in the right panel of Fig.~\ref{fig:EM}. The extracted constraints at 90\% C.L. read 
\begin{equation}
\begin{aligned}
\langle r_{\nu_e}^2\rangle =& (-64, -41) \, \, \text{and} \, \, (-7, 16)\, , \\  
\langle r_{\nu_\mu}^2\rangle =& (-69, -37) \, \, \text{and} \, \, (-10, 21)\, , \\  
\langle r_{\bar{\nu}_\mu}^2\rangle =& (-60,-43)\, \, \text{and} \, \, (-5, 12)\, , \\  
\end{aligned}
\end{equation}
in units of $~10^{-32}\mathrm{cm^2}$. Note that, comparing with the corresponding results extracted from the CsI case, the precision of the new determinations derived
from the new LAr measurement are significantly improved since now the allowed regions are narrower, and appear as two separate intervals.

At this point, and as we did for the NSI case, we have performed a combined analysis 
allowing for several parameters being non-zero at a time. In particular, we have chosen to probe the 
parameter space of neutrino magnetic moments $(\mu_{\nu_\alpha}, \mu_{\nu_\beta})$ and neutrino charge radii $(\langle r^2_{\nu_\alpha}\rangle, \langle r^2_{\nu_\beta}\rangle)$, by considering the simultaneous presence of two of them, i.e. $\alpha\neq\beta$.
The corresponding results are presented in Fig.~\ref{fig:comb-EM}, where we also compare with those derived from the 2017 data release of the COHERENT-CsI measurement. The  improvement obtained with the most recent data is more than evident.

\subsection{Light Mediators}

\begin{figure}[ht]
\includegraphics[width= \textwidth]{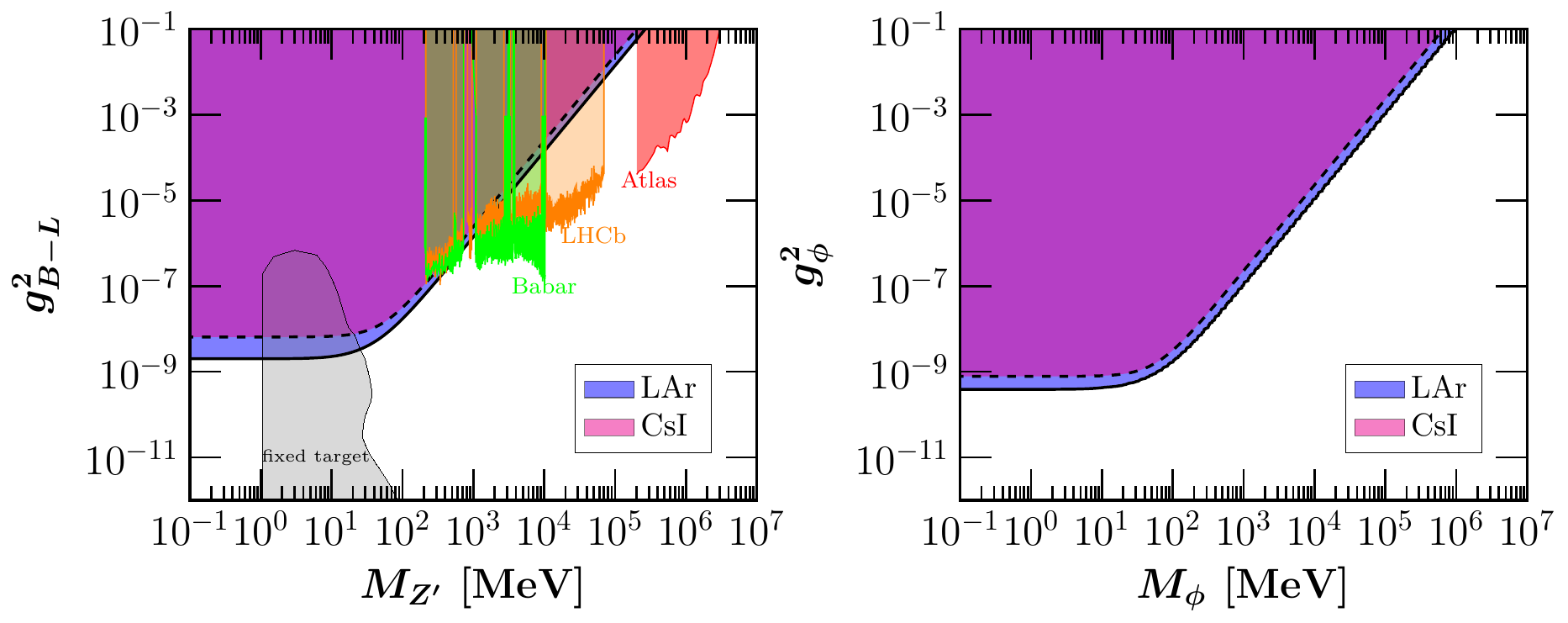}
\caption{Excluded region at  90\% C.L in the parameter space $(M_{Z^\prime},g_{B-L}^2)$ for the vector mediator scenario (left) and  $(M_{\Phi},g_{\Phi}^2)$ for the scalar mediator scenario (right), from the analysis of the recent LAr data. In both cases, a comparison is given with the CsI data. }
\label{fig:med}
\end{figure}

It has been recently shown~\cite{Cerdeno:2016sfi} that low energy scattering experiments are favorable facilities for probing the existence of light mediators, of both vector~\cite{Arcadi:2019uif} and scalar~\cite{Farzan:2018gtr} type. Indeed, by focusing on the $Z^\prime$ predicted within the context of string-inspired $E_6$ and Left-Right symmetries, the authors of Ref.~\cite{Miranda:2020zji} explored the potential of low-energy measurements at CEvNS experiments, concluding that complimentary information to high-energy collider searches can be achieved.  

Here, we focus on simplified  $U(1)^\prime$ scenarios with an additional vector $Z^\prime$ or a scalar $\phi$ boson that arise from the generic Lagrangians~\cite{Bertuzzo:2017tuf}
\begin{equation}
\begin{aligned}
{\cal L}_{\mathrm{vector}} = &  Z^{\prime}_\mu \left(g_{Z^\prime}^{qV} \bar{q} \gamma^\mu q + g_{Z^\prime}^{\nu V} \bar{\nu}_L \gamma^\mu \nu_L\right) \, , \\
{\cal L}_{\mathrm{scalar}} = & \phi \left(g_{\phi}^{q S} \bar{q} q + g_{\phi}^{\nu S} \bar{\nu}_R \nu_L + \mathrm{H.c.}\right) \, , 
\end{aligned}
\label{lagr:z-prime}
\end{equation}
with $M_{Z^\prime}$ and $M_\phi$ being the mass of the vector and scalar mediators, whereas $g_{Z^\prime}^{fV}$ and $g_{\phi}^{f S}$ are the respective vector and scalar couplings to the fermion $f=u,d,\nu$. For the case of a $Z^\prime$ mediator, there is an interference with the SM vector couplings, and the corrections to the SM cross section are incorporated through the substitution $\mathcal{Q}_W^V \rightarrow \mathcal{Q}_V^{Z^\prime}$, i.e. replacing the SM weak charge with the $Z^\prime$ one in Eq.(\ref{eq:xsec-cevns}), as~\cite{Papoulias:2019txv}
\begin{equation}
\mathcal{Q}_V^{Z^\prime} = 
\mathcal{Q}_W^V + \frac{g_{Z^\prime}^{\nu V}}{\sqrt{2}G_F}\frac{\left(2 g^{uV}_{Z^\prime} + g^{dV}_{Z^\prime} \right)Z F_p(Q^2) +  \left(g^{uV}_{Z^\prime} + 2g^{dV}_{Z^\prime} \right) N F_n (Q^2)}{2 m_A T_A + M_{Z^\prime}^2}  \, .
\label{eq:G_z-prime}
\end{equation}

Turning to the case of a scalar boson mediating the CEvNS process, there is no interference and, then, the total cross section is given by $(d \sigma/dT_A)_{\mathrm{tot}}= (d\sigma/dT_A)_{\mathrm{SM}} + (d \sigma/dT_A)_{\mathrm{scalar}}$, where the scalar contribution to the cross section is expressed as
\begin{equation}
\left(\frac{d \sigma}{d T_A} \right)_{\mathrm{scalar}} = \frac{G_F^2 m_A^2}{ 4 \pi} \frac{g_\phi^{\nu S} \mathcal{Q}_\phi^2  \, T_A }{E_\nu^2 \left( 2 m_A T_A + M_\phi^2 \right)^2} \, ,
\label{eq:xsec-scalar}
\end{equation}
and the scalar charge is defined as~\cite{AristizabalSierra:2019ykk}
\begin{equation}
\mathcal{Q}_\phi=Z F_p (Q^2)\sum_{q=u,d}g_\phi^{qS}\frac{m_p}{m_q}f_{T_q}^p
       +
       N F_n (Q^2)\sum_{q=u,d}g_\phi^{qS}\frac{m_n}{m_q}f_{T_q}^n \, .
\end{equation}
In the latter expression, the scalar charge is expressed in terms of the hadronic form factors $f_{T_q}^{q}$, obtained from chiral perturbation theory (see Ref.~\cite{Ellis:2018dmb}),
although here we use the updated values from~\cite{Hoferichter:2015dsa}
\begin{equation}
\begin{aligned}
  f_{T_u}^p&=(20.8\pm 1.5)\times 10^{-3}\ ,\;&
  f_{T_d}^p=(41.1\pm 2.8)\times 10^{-3}\ ,
  \nonumber\\
  f_{T_u}^n&=(18.9\pm 1.4)\times 10^{-3}\ ,\;&
  f_{T_d}^n=(45.1\pm 2.7)\times 10^{-3}\ .
\end{aligned}
\label{eq:ftqn}
\end{equation}

At this point, we should note that, for simplicity, we consider
  universal quark couplings for both vector and scalar cases,
  i.e. $g_{Z^\prime}^{uV}=g_{Z^\prime}^{dV}$ and
  $g_\phi^{uS}=g_\phi^{dS}$. Therefore, our sensitivity analysis will
  refer to the corresponding squared couplings entering in
  Eqs.~(\ref{eq:G_z-prime}) and (\ref{eq:xsec-scalar}) from the product
  of neutrino and quark couplings,
  i.e. $g_{Z^\prime}^2=g_{Z^\prime}^{qV}g_{Z^\prime}^{\nu V}$ and
  $g_{\phi}^2=g_{\phi}^{qS}g_{\phi}^{\nu S}$. Here, we find it interesting to focus on the $U(1)_{B-L}$ extension of the \sm where $g_{Z^\prime}^{qV}= -g_{Z^\prime}^{\nu V}/3$~\cite{Billard:2018jnl}. Using the latest data
  from the CENNS-10 measurement, we performed a combined analysis by
  varying simultaneously the vector (scalar) coupling and the
  corresponding vector (scalar) mediator mass. The corresponding excluded
  regions are illustrated in Fig.~\ref{fig:med} and compared with the COHERENT-CsI case. For the vector mediator scenario, our results are also compared with existing limits placed by 
dielectron resonances at ATLAS~\cite{Aaboud:2016cth}, constraints from electron beam-dump fixed target experiments~\cite{Harnik:2012ni,Ilten:2018crw} as well as with
constraints from Dark Photon searches at BaBar~\cite{Lees:2014xha,Lees:2017lec} and LHCb~\cite{Aaij:2017rft}. One sees that CEvNS searches are clearly complementary to the latter ones, excluding a large part of the available parameter space. Notice the slight improvement found with respect to the first COHERENT-CsI measurement.

\subsection{Robustness of the constraints}
\label{sec:robustn-constr}

Nuclear physics uncertainties place important limitations on the attainable sensitivities to physics observables extracted from coherent neutrino elastic scattering experiments.
  Indeed, as emphasized in Refs.~\cite{Papoulias:2019lfi, AristizabalSierra:2019zmy,Canas:2019fjw}, these may lead to a miss-interpretation of the relevant constraints derived from CEvNS measurements.

\begin{figure}[t]
\includegraphics[width= 0.48 \textwidth]{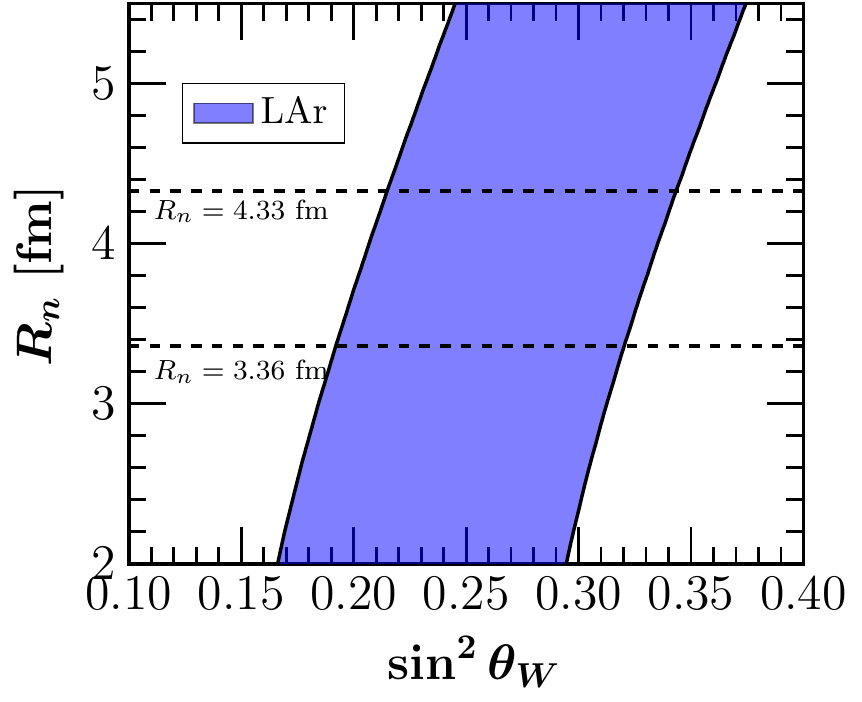}
\includegraphics[width= 0.48 \textwidth]{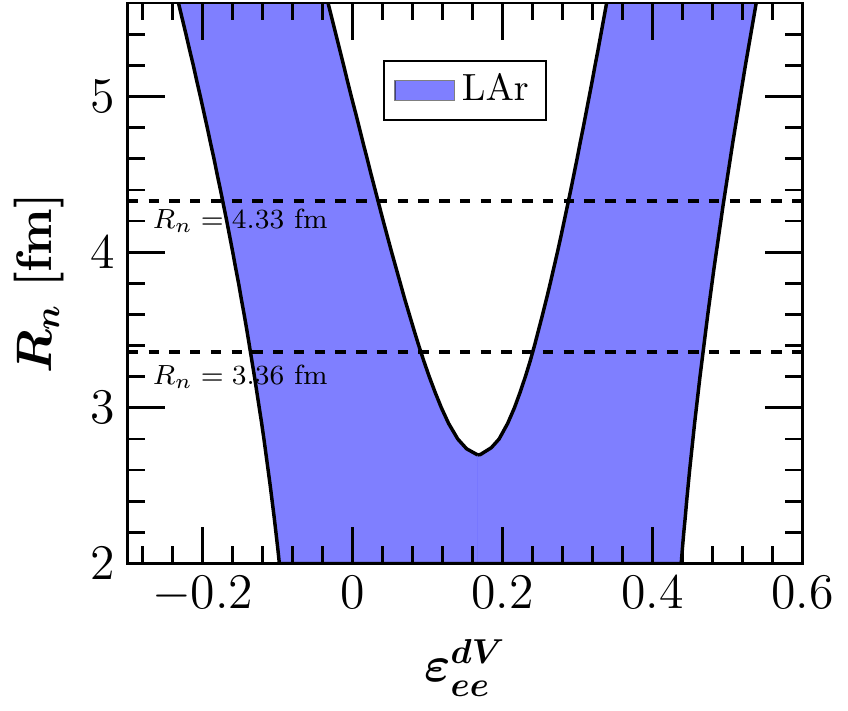}
\caption{Allowed region in the parameter space of $(\sin^2 \theta_W, R_n)$  (left) and  $(\epsilon_{ee}^{dV}, R_n)$ (right) planes at 90\% C.L. For comparison, the neutron rms radii $R_n=3.36~\text{fm}$ (fixed value) and $R_n=4.33~\text{fm}$ [upper limit  in Eq.(\ref{eq:Rn-limit})] are   indicated with horizontal dashed lines.}
\label{fig:comb-robustness}
\end{figure}

Therefore, before closing our present analysis, we find it useful to devote a separate paragraph in order to discuss the robustness of the constraints we have obtained with regards to the nuclear form factor. To this purpose, we performed a combined analysis of the weak mixing angle and the  neutron rms radius  simultaneously; we also performed a similar analysis for the case of a NSI parameter characterizing new physics. 
While many such combinations are possible, as a concrete example in Fig.~\ref{fig:comb-robustness} we show the allowed regions in the
  $(\sin^2 \theta_W, R_n)$ and $(\epsilon_{ee}^{dV}, R_n)$ planes at 90\% C.L.
As expected, in the left panel, one sees how the 90\% C.L. determination of the weak mixing angle has a larger relative error for free $R_n$, in comparison to that of Eq.(\ref{eq:sin2w-limit}) obtained with the fixed value $R_n=3.36$~fm. We find $\delta s^2_W (\text{free}~R_n)=0.128$ vs. $\delta s^2_W(\text{fixed}~R_n)=0.097$, where $\delta s_W^2$  corresponds to the width of the 90\% C.L. band.
On the other hand, concerning new physics, we show in the right panel the allowed region on the NSI parameter $\varepsilon_{ee}^{dV}$ for different values of the neutron rms radius $R_n$.
One sees that, by using a free rms neutron radius $R_n$, the 90\% C.L. leads to two disjoint ranges (for $R_n>2.7$~fm) and a reduced sensitivity compared to the results obtained from the analysis with a fixed rms radius shown in Fig.~\ref{fig:NSI}.  
In both cases, it becomes evident that the limitations imposed due to the nuclear physics uncertainties must be treated with special care. In fact, this may require realistic nuclear structure calculations~\cite{Papoulias:2013gha,Papoulias:2019lfi}.

\section{Conclusions}
\label{sec:conclusions}

We have analyzed the recent results of the CENNS-10 detector subsystem of the COHERENT collaboration that led to the first detection of CEvNS on LAr.
Through a dedicated statistical analysis, taking into account the available information from Ref.~\cite{Akimov:2020pdx}, we have shown that this new measurement typically leads
to improved sensitivities with respect to the first COHERENT-CsI measurement in 2017.
Specifically, we have presented an improved determination of the weak mixing angle, as well as the first ever determination of the $^{40}$Ar neutron rms radius.
Turning to new physics,  we have derived the constraints on non-universal as well as flavor-changing NSI imposed by this new data release.
Moreover, concerning neutrino electromagnetic properties, we have found only minor improvement of the sensitivity to neutrino magnetic moments.
In contrast, we have found a positive indication for finite neutrino-charge radii.
We have shown that the new CENNS-10 data provides somewhat better sensitivities on simplified scenarios involving new light mediators when compared to COHERENT-CsI data,
  discussing also the complementarity to high energy experiments.
Finally, we have explored the impact of nuclear physics uncertainties and discussed the robustness of our results.

%%%%%%%%%%%%%%%%%%%%%%%%%%%%%%%%%%%%%%%%%%%%%%%%%%%%%%%%%%%%%%%%%%%%

\begin{acknowledgments}
The authors acknowledge Carlo Giunti for useful comments.
This work is supported by the Spanish grants FPA2017-85216-P
(AEI/FEDER, UE), PROMETEO/2018/165 (Generalitat Valenciana) and the Spanish Red Consolider MultiDark FPA2017-90566-REDC, and by
CONACYT-Mexico under grant A1-S-23238. OGM has been supported by SNI (Sistema Nacional de Investigadores). The work of DKP is co-financed by Greece and the European Union (European Social Fund- ESF) through the Operational Programme <<Human Resources Development, Education and Lifelong Learning>> in the context of the project ``Reinforcement of Postdoctoral Researchers - 2nd Cycle" (MIS-5033021), implemented by the State Scholarships Foundation (IKY). MT acknowledges financial
support from MINECO through the Ram\'{o}n y Cajal contract
RYC-2013-12438.

\end{acknowledgments}

%%%%%%%%%%%%%%%%%%%%%%%%%%%%%%%%%%%%%%%%%%%%%%%%%%%%%%%%%%%%%%%%%%%%

%
\bibliographystyle{utphys}
\bibliography{bibliography}

\end{document}